\documentclass[runningheads]{llncs}
\usepackage{graphicx}
\usepackage{amsmath}
\usepackage{amssymb}
\usepackage[utf8]{inputenc}
\usepackage{todonotes}
\usepackage{siunitx} 
\usepackage[hidelinks]{hyperref}
\usepackage[labelformat=simple]{subcaption}
\usepackage{booktabs}
\usepackage{multirow}
\usepackage{tabu}
\usepackage[dvipsnames]{colortbl}
\usepackage{textcomp}
\usepackage{xcolor}
\usepackage[misc]{ifsym}

\makeatletter
\newcommand\footnoteref[1]{\protected@xdef\@thefnmark{\ref{#1}}\@footnotemark}
\makeatother

\begin{document}
\title{An Analysis by Synthesis Method that Allows Accurate Spatial Modeling of Thickness of Cortical Bone from Clinical QCT}
\titlerunning{A Method for Spatial Modeling of Cortical Thickness from Clinical QCT}
\author{\href{https://orcid.org/0000-0003-3117-1569}{Stefan Reinhold}\inst{1}\href{mailto:sre@informatik.uni-kiel.de}{\textsuperscript{\Letter}}\and
\href{https://orcid.org/0000-0002-5595-5205}{Timo Damm}\inst{3}\and
\href{https://orcid.org/0000-0002-1657-7950}{Sebastian Büsse}\inst{2}\and
\href{https://orcid.org/0000-0001-9712-7953}{Stanislav N. Gorb}\inst{2}\and
\href{https://orcid.org/0000-0003-3539-8955}{Claus-C. Glüer}\inst{3}\and
\href{https://orcid.org/0000-0003-4398-1569}{Reinhard Koch}\inst{1}}
%
\authorrunning{S. Reinhold et al.}
\institute{Department of Computer Science, Kiel University, Kiel, Germany\\\email{\{\underline{sre},rk\}@informatik.uni-kiel.de}\and
Functional Morphology and Biomechanics, Institute of Zoology, Kiel University, Kiel, Germany\\
\email{\{sbuesse,sgorb\}@zoologie.uni-kiel.de}\and
Section Biomedical Imaging, Molecular Imaging North Competence Center (MOIN CC), Department of Radiology and Neuroradiology, University Medical Center Schleswig-Holstein (UKSH), Kiel University, Kiel, Germany\\
\email{\{timo.damm,glueer\}@rad.uni-kiel.de}}
\maketitle
\begin{abstract}
Osteoporosis is a skeletal disorder that leads to increased fracture risk due to decreased strength of cortical and trabecular bone.
Even with state-of-the-art non-invasive assessment methods there is still a high underdiagnosis rate.
Quantitative computed tomography (QCT) permits the selective analysis of cortical bone, however the low spatial resolution of clinical QCT leads to an overestimation of the thickness of cortical bone (Ct.Th) and bone strength.

We propose a novel, model based, fully automatic image analysis method that allows accurate spatial modeling of the thickness distribution of cortical bone from clinical QCT.
In an analysis-by-synthesis (AbS) fashion a stochastic scan is synthesized from a probabilistic bone model, the optimal model parameters are estimated using a maximum a-posteriori approach.
By exploiting the different characteristics of in-plane and out-of-plane point spread functions of CT scanners the proposed method is able assess the spatial distribution of cortical thickness.

The method was evaluated on eleven cadaveric human vertebrae, scanned by clinical QCT and analyzed using standard methods and AbS, both compared to high resolution peripheral QCT (HR-pQCT) as gold standard.
While standard QCT based measurements overestimated Ct.Th. by 560\% and did not show significant correlation with the gold standard ($r^2 = 0.20,\, p = 0.169$) the proposed method eliminated the overestimation and showed a significant tight correlation with the gold standard ($r^2 = 0.98,\, p < 0.0001$) a root mean square error below 10\%.

\end{abstract}
\setcounter{footnote}{0} 

\section{Introduction}

\begin{figure}[tb]
	\includegraphics[width=12cm]{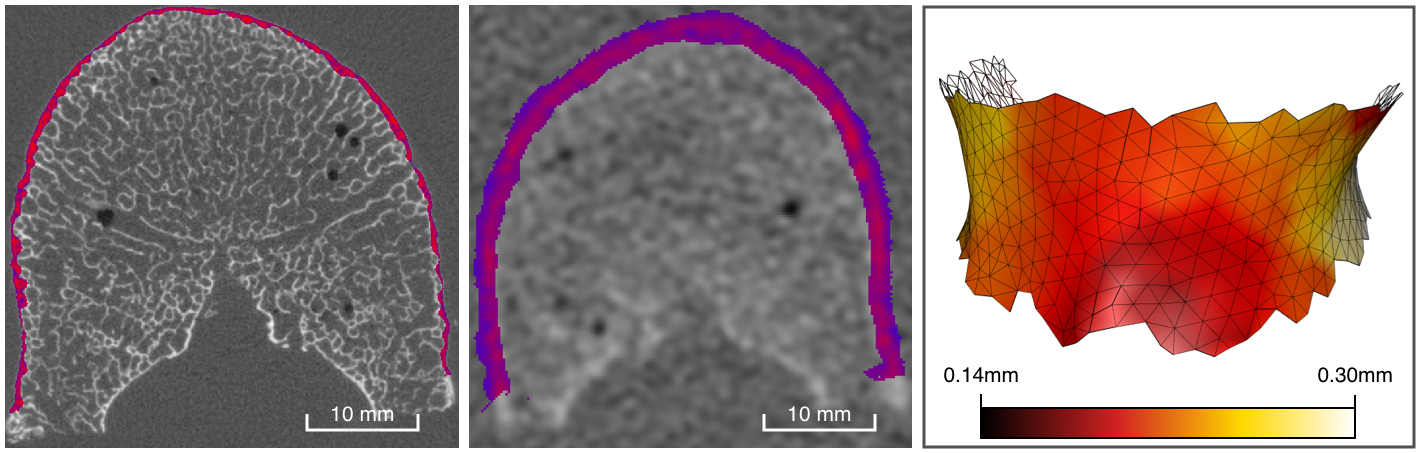}
	\caption[Comparison XCT vs. QCT]{Comparison of HR-pQCT (left) and standard QCT (center): axial slice\footnotemark of a vertebral body. The apparent cortical bone is highlighted in red. The sponge like structure of trabecular bone is clearly visible in the left image. Right: vertical cortex mesh with color coded thickness estimated by our AbS algorithm.}
	\label{fig:XCT-vs-QCT}
\end{figure}

The world health organization (WHO) estimates the life time risk of a osteoporotic fracture at 30-40\% \cite{who2004}. Even though non-invasive assessment methods are widely available, there is still a very high underdiagnosis rate\cite{smith2004treatment}.
Current osteoporosis diagnosis is based on the assessment of trabecular bone mineral density (BMD).
However, recent studies \cite{ESWARAN2007,YAMADA2019} show that the thin cortical shell contributes at least 50\% to overall bone strength and
that there is a strong correlation between the thickness of vertebral cortical bone (Ct.Th) and failure load estimated by finite element analysis (FEA).
While quantitative computed tomography (QCT) permits the selective analysis of cortical bone, the limited spatial resolution (pixel size \num{300}-\num{500}\si{\micro\metre}, slice width \num{1}-\num{3}\si{\milli\metre}) of clinical QCT  prevents accurate assessment.
The thickness of the thin cortical shell (\num{150}-\num{400}\si{\micro\metre}) is highly overestimated by current methods.
Figure \ref{fig:XCT-vs-QCT} shows a comparison of a scan taken with high resolution peripheral QCT (HR-pQCT) and clinical QCT protocols.
The amount of overestimation in the clinical QCT scan is clearly visible.

\footnotetext{Due to different orientation of the specimen the two slices appear flipped.}

\subsubsection{Related work}
BMD is defined as the volumetric density of calcium hydroxyapatite (CaHA). 
However, typical BMD measurements also include surrounding soft tissue and marrow.
\emph{Tissue mineral density} (TMD) is the density of the calcified bone, excluding soft tissue and lies around \num{1200}\si{\milli\gram\per\cubic\centi\metre} for fully mineralized bone\cite{Kazakia2008,KAZAKIA2011,laib1998,laval1983}.
The standard method to assess Ct.Th is to apply a maximum sphere approach to a segmented scan\cite{Hildebrand1997}.
Since it solely depends on the cortex as apparent in the reconstructed volume, the thickness measurements from this standard method is denoted as \emph{apparent Ct.Th} (aCt.Th) throughout this article.
As aCt.Th is highly affected by the spatial resolution of the reconstructed volume Ct.Th is tremendously overestimated in clinical QCT while cortical BMD (Ct.BMD) underestimates cortical TMD.
One way to account for the overestimation is to combine thickness with density measurements into \emph{density weighted Ct.Th} (wCt.Th) by multiplying aCt.Th with Ct.BMD normalized by a TMD of \num{1200}\si{\milli\gram\per\cubic\centi\metre}, however this method is still limited by the spatial resolution and the quality of the segmentation.

Prevrhal et al. \cite{Prevrhal2003,Prevrhal1999} introduced a method for improved assessment of Ct.Th by using analytical models to describe the blurring of the imaging system.
However the improvement vanishes for thin cortices.
Hangartner et al. proposed an iterative mathematical model to correct aCt.Th and Ct.BMD measurements based on per scanner density-versus-width curves for peripheral CT, yielding low error rates in phantom experiments for cortices thicker than \num{0.5}\si{\milli\metre}.
Other improvements based on star-line tracing \cite{Liu2014} or fuzzy distance transform \cite{Li2015} exist for HR-pQCT of tibia.
However, those method are not applicable to vertebrae where HR-pQCT is not an option in a clinical setting.
Treece et al. \cite{TREECE2010276,TREECE2012952,TREECE2015249} describe a method to correct thickness measurement from clinical QCT of the femoral cortex.
Damm et al. \cite{DAMM2019194} propose an \emph{Iterative Convolution OptimizatioN} (ICON) method.
They showed that \emph{deconvolved Ct.Th} (dcCt.Th) can reduce the overestimation of Ct.Th to 20\% in high resolution QCT (HR-QCT) and increase the correlation between clinical QCT and HR-pQCT thickness measurements.

Recently, Reinhold et al. \cite{Reinhold2019} proposed a method to identify the center of the vertebral cortical bone with sub-voxel accuracy.
They used an analysis-by-synthesis (AbS) approach to fit a geometrical model of the cortical shell to the input scan.

\subsubsection{Our Contribution}
We propose a fully automatic AbS based algorithm that allows the accurate spatial modeling of the thickness distribution of cortical bone from clinical QCT.
Starting from an approximate surface along the center of the cortical bone, the bone thicknesses and densities are modeled as latent variables of a stochastic measurement process.
We exploit the different characteristics of the in-plane and out-out-plane point spread functions (PSF) of CT scanners in our optimization procedure.
For that we propose a novel analytical approximation of the in-plane PSF that meets the required accuracy.
To estimate the a-posteriori distributions of the latent model, a Monte Carlo expectation maximization (MCEM) scheme, tailored to our requirements is proposed.
The result is an accurate spatial model of the cortical bone permitting
the analysis of intra bone variations of Ct.Th which has the potential in improving future FEA and osteoporosis diagnosis in general.

In chapter \ref{sec:AlgorithmOverview} an overview of the proposed algorithm is given.
It is evaluated in an ex-vivo experiment, described in chapter \ref{sec:Methods}; we compare our approach with standard methods.
After presenting and discussing the results in chapter \ref{sec:results}, we conclude this articles in chapter \ref{sec:conclusion}.

\section{Algorithm Overview}
\label{sec:AlgorithmOverview}

The aim of our algorithm is to estimate the parameters of a three dimensional cortex model from a clinical QCT scan (fig. \ref{fig:XCT-vs-QCT} right).
Because of the low spatial resolution the imaging process cannot be reversed.
Instead, a synthetic version of a reconstructed volume is generated from the model; the synthesis results are compared with the input volume until the best model parameters are found.
To make this AbS process feasible, simplifications and approximations must be made.
First of all, it is assumed that the cortical bone is a dense plate with varying thickness and density, modeled as a triangle mesh along the cortex center with per-vertex thickness and density properties.
Although the trabecular region consists of trabecles surrounded by bone marrow (fig. \ref{fig:XCT-vs-QCT} left), it is modeled as an area with homogeneous density.
This choice is justified by the fact that the synthesis results for discretely modeled trabecles are indistinguishable from a homogeneous density distribution for low resolutions (fig. \ref{fig:XCT-vs-QCT} center).
Since simulating a complete CT reconstruction in each synthesis step is computationally expensive, the imaging system is approximated by a blurring with an in-plane and an out-out-plane PSF.
Furthermore, the complete synthesis of a volume is avoided by using sparse synthesis where only one dimensional profiles perpendicular through the cortex are synthesized.

\subsubsection{Bone Model}
\begin{figure}[tb]
	\begin{subfigure}[c]{0.5\textwidth}
		\includegraphics[width=\textwidth, height=3.5cm]{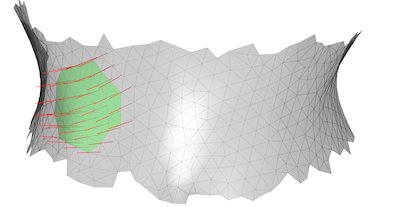}
	\end{subfigure}
	\begin{subfigure}[c]{0.5\textwidth}
		\includegraphics[width=\textwidth,height=3.5cm]{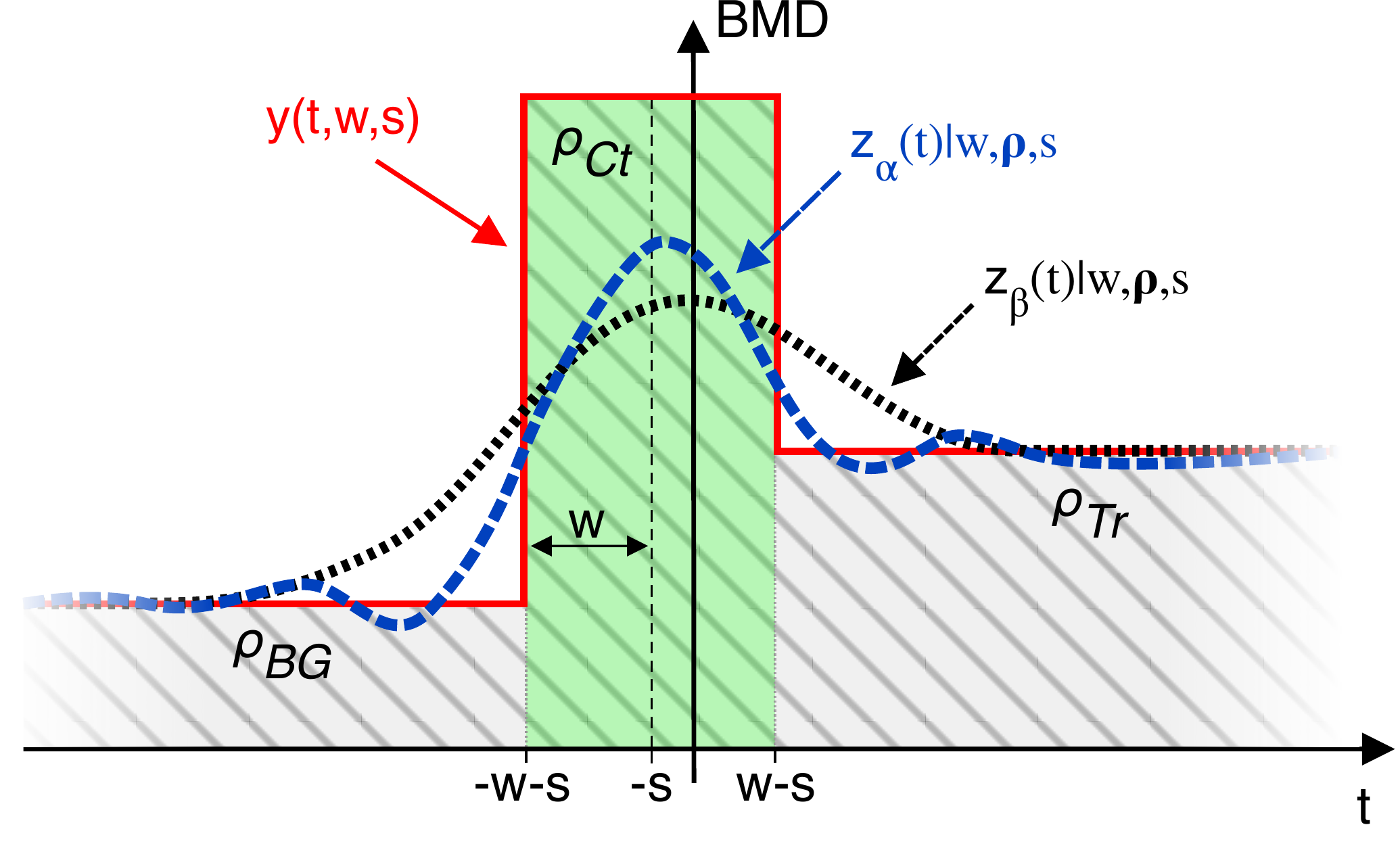}
	\end{subfigure}
	\caption{Left: Cortex surface mesh with local patch (green); profiles (red lines) used to sample the input volume. Right: Schematic bone model: the cortex (green) center is slightly shifted. Two possible measurements of the same underlying signal under different angles are depicted (dashed / dotted).}
	\label{fig:BoneAndMeasurementModel}
\end{figure}
The bone model is an extension of the one used in \cite{Reinhold2019}. 
Following their notation, the density of the background (e.g. surrounding soft tissue), the cortical and the trabecular bone is modeled as a multivariate gaussian latent random vector $\vec{\rho} = (\rho_{BG}, \rho_{Ct}, \rho_{Tr})^{\operatorname{T}} \sim \mathcal{N}(\vec{\mu}_{\vec{\rho}}, \Sigma_{\vec{\rho}})$ and the half cortex width $w$ is modeled as a log normally distributed latent random variable with $\log w \sim \mathcal{N}(\mu_w, \sigma_w^2)$.
Due to the shape constraints in the cortex identification process, there might be a small offset from any point on the estimated surface to the real cortex center.
To account for this small but unknown offset $s \sim \mathcal{N}(\mu_s, \sigma_s^2)$, we add it as another latent variable to the profile process:
\begin{equation}
	\label{eq:boneProcess}
	y(t, w, s) = \Phi(t - s, w) \cdot \vec{\rho},
\end{equation}
where $
	\Phi(t, w) = \left[
		1 - \operatorname{H}(t + w),\,
		\operatorname{H}(t + w) - \operatorname{H}(t - w),\,
		\operatorname{H}(t - w)
		\right], \nonumber
		$
$t$ is the position along the profile and $\operatorname{H}$ is the Heaviside step function (see fig. \ref{fig:BoneAndMeasurementModel} (right) for an example).
To simplify the model, we assume that all latent variables $\vec{\rho}$, $w$, $s$ are independent and $\Sigma_{\vec{\rho}}$ is diagonal.
From literature and $\mu$CT experiments we have a basic understanding of the distribution of those parameters.
Therefore we define a joint weakly informative Normal-Inverse-$\chi^2$ (NI$\chi^2$) prior on the parameter vector $(\vec{\mu}_{\vec{\rho}}^{\operatorname{T}}, \mu_w, \mu_s, \operatorname{diag}(\Sigma_{\vec{\rho}})^{\operatorname{T}}, \sigma_w^2, \sigma_s^2)^{\operatorname{T}}$.

\subsubsection{Measurement Model}
As in \cite{Reinhold2019}, the imaging process is approximated by a blurring with an in-plane and and out-of-plane PSF.
Both PSFs can be combined into a one dimensional, angle dependent PSF $g_{\alpha}$ along a profile.
The influence of each PSF varies with the angle of the profile with the z-axis.
Profiles from regions with equal latent distributions but different angles will therefore have slightly different synthesis results (fig. \ref{fig:BoneAndMeasurementModel} right).
This is the key observation that leads to our algorithm.
However, in contrast to previous works \cite{Reinhold2019,DAMM2019194,TREECE2010276,TREECE2012952,TREECE2015249}, a convenient gaussian approximation for the in-plane PSF is not sufficient anymore.
Using a normalized symmetric sum of gaussian in the Fourier domain, the PSF can be approximated with arbitrary precision. After inverse Fourier transform the in-plane PSF states:
\begin{equation}
	g_{\text{ip}}(t)
	= \left(
		2 \sum_{k=1}^{N_{\text{ip}}} a_k \exp \left( {\scriptstyle
		- \frac{b_k^2}{2c_k^2}
		} \right)
	\right)^{-1} \sum_{k=1}^{N_{\text{ip}}} a_k \xi_k(t),
\end{equation}
where
$\xi_k(t) = \sqrt{2 \pi} c_k \left[ \exp\left( {\scriptstyle -2 \pi t (\pi c_k^2 t - \imath b_k)} \right)
	+ \exp\left( {\scriptstyle
	- 2 \pi t (\pi c_k^2 + \imath b_k)
	}\right) \right]$ and $a_k, b_k, c_k \in \mathbb{R},\, c_k > 0$ are obtained by fitting the positive part of $\mathcal{F}\{g_{\text{ip}}\}$ to an empirically measured MTF.
Given the combined PSF $g_{\alpha}$ the stochastic measurement process can be formulated:
\begin{equation}
	z_{\alpha}(t)|\vec{\rho}, w, s = \Psi_{\alpha}(t, w, s) \cdot \vec{\rho} + g_{\alpha}(t) \ast \epsilon(t) + \xi(t),
\end{equation}
where $\Psi_{\alpha}(t, w, s) = \Phi(t - s, w) \ast g_{\alpha}(t)$, $\epsilon(t) \sim \mathcal{N}(0, \sigma_{\epsilon}^2)$ is a gaussian white noise process simulating the measurement noise, $\xi(t) \sim \mathcal{N}(0, \sigma_{\xi}^2)$ is a gaussian noise process accounting for model errors and the notation $z_\alpha(t)|\vec{\rho}, w, s$ denotes that the stochastic process $z_\alpha(t)$ is conditioned on the random variables $\vec{\rho}$, $w$ and $s$.

\subsubsection{Optimization}
Let $Z$ be a set of profiles sampled from the input volume, perpendicular to the cortex surface (cf. fig. \ref{fig:BoneAndMeasurementModel} left).
Each profile in $Z$ is assumed to be a realization of the stochastic process $z_\alpha(t)|\vec{x},\vec{\theta}$,
where $\vec{x} = (\ln w, \vec{\rho}^{\operatorname{T}}, \vec{s}^{\operatorname{T}})^{\operatorname{T}}$ is the coalesced random vector and 
$\vec{\theta} = (\mu_w, \sigma_w, \vec{\mu}_{\rho}^{\operatorname{T}},\vec{\sigma}_{\rho}^{\operatorname{T}}, \vec{\mu}_{s}^{\operatorname{T}}, \vec{\sigma}_{s}^{\operatorname{T}})^{\operatorname{T}}\sim \operatorname{NI\chi^2}(\vec{\theta}_0)$ is the vector of its distribution parameters.
Our goal is to find the parameters $\vec{\theta}^\star$ that maximize the posterior density of $\vec{\theta}$ given $Z$.
However, the posterior density contains an intractable integral over $\mathbb{R}^{N+4}$.
Using an expectation maximization (EM) scheme, only the following lower bound needs to be maximized:
\begin{equation}
	\ln p(\vec{\theta}|Z, \vec{\theta}_0) \geq \int p(\vec{x}|Z, \vec{\theta}^{[i]}) \ln \left( p(\vec{x} | \vec{\theta}) p(\vec{\theta} | \vec{\theta}_0) \right) d\vec{x}.\label{eq:LowerBoundLogPosteriorLLExplicit}
\end{equation}
Using Monte Carlo integration, \eqref{eq:LowerBoundLogPosteriorLLExplicit} can be approximated as
\begin{equation}
	\int p(\vec{x} | Z, \vec{\theta}^{[i]}) \ln \left( p(\vec{x} | \vec{\theta}) p(\vec{\theta} | \vec{\theta}_0) \right) d\vec{x} \approx \frac{1}{K} \sum_{k=1}^{K} \gamma_i \ln p(\vec{\theta} | \vec{x}_k, \vec{\theta}_0),\label{eq:approxPosterior}
\end{equation}
with $\vec{x}_k \sim q$, $\gamma_i = p(\vec{x}_k|Z, \vec{\theta}^{[i]})(q(\vec{x}_k))^{-1}$.
Since the posterior $p(\vec{\theta} | \vec{x}_k, \vec{\theta}_0)$ is NI$\chi^2$, eq. \eqref{eq:approxPosterior} can be maximized in closed form\cite{Lee2012}, yielding the estimate $\vec{\theta}^{[i+1]}$ for the next iteration.
As $p(\vec{x}_k|Z, \vec{\theta}^{[i]})$ is intractable, we use an iterative adaptive multiple importance sampling scheme based on the work of El-Laham et al. \cite{ElLaham2018} to approximate it.
To improve the convergence properties of the algorithm, an ascend-based MCEM scheme \cite{caffo2005} is utilized.
To avoid local maxima, we start with a small sample size $K$, increase it dependent on the amount of improvement that was made in the last few iterations ensuring convergence.
The algorithm stops when the estimated upper bound of the log likelihood improvement falls below a given threshold.

\section{Materials and Methods}
\label{sec:Methods}
Eleven excised human vertebrae, obtained from the anatomical institute of our institution were examined. 
Ethics were approved by the responsible ethics review committee.
The vertebrae were embedded into a body phantom and scanned\footnote{\label{note:parameters}A complete list of all parameters can be found in the supplement.} on a CT scanner with a clinical QCT (resolution  \num{0.234}\texttimes \num{0.234}\texttimes \num{1}\si{\milli\metre}) and a HR-pQCT protocol (isotropic resolution \num{0.082}\si{\milli\metre}).
Hounsfield Units (HU) were calibrated to \si{\milli\gram} CaHA \si{\per\cubic\centi\metre} using a QRM CT calibration phantom.
Several slices of cortical bone were acquired from a non-embedded excised vertebra for detailed analysis in a high-resolution \textmu CT scanner\footnoteref{note:parameters} (isotropic resolution \num{1.73}\si{\micro\metre}).

\begin{figure}[tb]
	\centering
	\includegraphics[width=12cm,height=2.62cm]{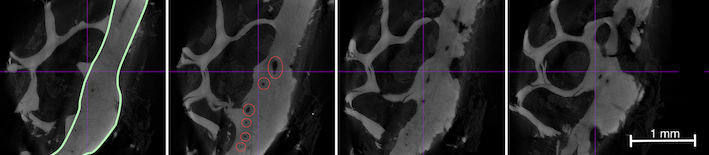}
	\caption{High-resolution \textmu CT scans of four different cortical bone slices. The cortex has been outlined in green on the first slice to highlight the variability of the cortical thickness. Pores have been marked in red on the second slice.}
	\label{fig:ultraCT}
\end{figure}

The cortices of the clinical QCT scans were identified using the method of Reinhold et al \cite{Reinhold2019}.
The resulting triangle meshes were used to generate a voxel based segmentation and as starting point for the proposed AbS method.
The HR-pQCT scans were segmented using the dual thresholding method of Buie et al \cite{BUIE2007505}. 
An established in-house software tool was used to assess standard aCt.Th and wCt.Th for the vertical cortex region of the segmented scans, QCT and HR-pQCT.
The vertical cortex region of the surface mesh was divided into 48 quasi randomly placed, partially overlapping patches\footnote{With 48 patches the vertical cortex was completely covered with minimal overlap.}.
For each patch between 11 and 51 profiles were sampled from the QCT scans, excluding profiles outside the target VOI.
Those profiles were used as the input for the proposed method and
cortical thickness distributions were estimated for each scan and patch using our AbS method\footnoteref{note:parameters}.
For comparison with other methods a per-specimen average was computed by merging all patch distributions into a single distribution, adjusting for overlaps\footnote{A single big patch is not feasible since the dimension of the latent space linearly increases with the patch size requiring an exponentially increasing number of samples.}.
To gather per patch reference data, the statistics of HR-pQCT wCt.Th in cylindrical regions approximately reflecting the corresponding surface patches of the QCT meshes were computed.

All statistical calculations were performed using the R \cite{R} programming language.
Levels of probability $p < 0.05$ were considered significant.

\section{Results and Discussion}
\label{sec:results}

\begin{table}[tb]

\caption{\label{tab:meanDeviation}Mean deviation from and correlation with gold standard by method.}
\centering
\begin{tabu} to \linewidth {>{\raggedright\arraybackslash}p{4.5cm}>{\centering\arraybackslash}p{1.75cm}>{\centering\arraybackslash}p{1.75cm}>{\centering\arraybackslash}p{1.75cm}>{\centering\arraybackslash}p{1.75cm}}
\toprule
\multicolumn{1}{c}{ } & \multicolumn{1}{c}{Mean\textpm SD} & \multicolumn{2}{c}{Mean\textpm SD deviation} & \multicolumn{1}{c}{Correlation} \\
 & [mm] & [mm] & [\%] & [$r^2$]\\
\midrule
\addlinespace[0.3em]
\multicolumn{5}{l}{\textit{QCT}}\\
\rowcolor[HTML]{b6f6b5}  \hspace{1em}Proposed\textsuperscript{\dag} & 0.29 \textpm 0.08 & -0.05 \textpm 0.03 & -15 \textpm  8 & $0.98^{***}$\\
\hspace{1em}Standard (aCt.Th) & 2.24 \textpm 0.13 & 1.90 \textpm 0.13 & 560 \textpm 37 & $0.20^{n.s.}$\\
\hspace{1em}Density Weighted (wCt.Th) & 0.62 \textpm 0.11 & 0.28 \textpm 0.05 & 83 \textpm 14 & $0.82^{***}$\\
\hline
\addlinespace[0.3em]
\multicolumn{5}{l}{\textit{HR-pQCT}}\\
\hspace{1em}Standard (aCt.Th) & 0.69 \textpm 0.17 & 0.35 \textpm 0.06 & 103 \textpm 18 & $0.99^{***}$\\
\hspace{1em}Density Weighted (wCt.Th) & 0.34 \textpm 0.11 & -- & -- & --\\
\bottomrule
\multicolumn{5}{l}{\textsuperscript{n.s.} $p \geq 0.05$; *** $p < 0.001$; \textsuperscript{\dag} Evaluated on 48 patches per specimen}\\
\end{tabu}
\end{table}
The aim of our method is to spatially model the real thickness distribution of the cortical bone.
But even the resolution of HR-pQCT is not sufficient to assess Ct.Th directly.
As can be seen in the second last row of table \ref{tab:meanDeviation}, the average HR-pQCT based aCt.Th 
is \num{0.69} \textpm \num{0.17}\si{\milli\metre} which is clearly above the thickness range reported in literature\cite{Ritzel1997}.
For HR-pQCT, wCt.Th should be a good approximation of the real Ct.Th.
Figure \ref{fig:ultraCT} shows four representative slices of cortical bones obtained by high-resolution \textmu CT.
The cortical bone is visible as a compact structure with only few small pores.
Its thickness varies between locations; the span is similar to the one reported in literature\cite{Ritzel1997}.
We therefore choose HR-pQCT wCt.Th as the gold standard for the following evaluations.

\subsubsection{Average Cortical Thickness}
Table \ref{tab:meanDeviation} compares the mean deviation from gold standard for the proposed method, QCT based aCt.Th and wCt.Th.
Our AbS method completely eliminates the overestimation introduced by all other methods and tightly correlates with the gold standard; it is able to explain 98\% of the variance.
A Bland-Altman analysis\cite{Krouwer2008}, showed that a slight proportional error might be introduced but there is still a good agreement with the gold standard.
We would like to remark here that the proposed method was only evaluated at 48 patches not on the complete vertical cortex and the error might by induced by the different samples\footnote{Additional supporting tables and figures can be found in the supplement.}.
Nevertheless, our method reduces the mean deviation to less than 25\% of the in-plane pixel size and is able to tightly correlate Ct.Th estimated from clinical QCT scans to wCt.Th measurements from HR-pQCT.

\subsubsection{Intra-Bone Correlation Analysis}
Unlike other methods, our AbS method is able to model the spatial thickness distribution.
Figure \ref{fig:XCT-vs-QCT} (right) shows the resulting distribution for a representative vertebra.
Visualizations of the spatial distribution of the differences between our method and the gold standard can be found in figure 4 in the supplement.
Table \ref{tab:intraBoneCorrelation} depicts the correlation of the proposed method with the gold standard per specimen.
For all specimens there is a significant ($p < 0.001$) correlation able to explain 40\% - 89\% of the variance; the RMSE ranges from 6\% to 29\% but does not exceed \num{70}\si{\micro\metre} which is below 30\% of the in-plane and below 7\% of the out-of-plane resolution of the input scan.
Our method permits to estimate a complete three-dimensional model of the cortical bone: the initial surface mesh is augmented by local per vertex thickness estimates and can be transferred into a volumetric tetrahedral mesh.
Starting from there, a ready to use finite element model could be generated that accurately models spatial thickness variations.
Such a model should permit more accurate finite element analysis. 
\begin{table}[t]

\caption{\label{tab:intraBoneCorrelation}Intra-bone correlation between the proposed method and gold standard.}
\centering
\begin{tabu} to \linewidth {>{\raggedright\arraybackslash}p{5em}>{\raggedright\arraybackslash}p{3em}>{\raggedleft}X>{\raggedleft}X>{\raggedleft}X>{\raggedleft}X>{\raggedleft}X>{\raggedleft}X>{\raggedleft}X>{\raggedleft}X>{\raggedleft}X>{\raggedleft}X>{\raggedleft}X}
\toprule
 &  & V1 & V2 & V3 & V4 & V5 & V6 & V7 & V8 & V9 & V10 & V11\\
\midrule
Correlation & [$r^2$] & 0.69 & 0.74 & 0.76 & 0.42 & 0.75 & 0.49 & 0.54 & 0.89 & 0.60 & 0.83 & 0.40\\
\rowcolor[HTML]{b6f6b5}  RMSE & [mm] & 0.02 & 0.03 & 0.07 & 0.06 & 0.06 & 0.04 & 0.07 & 0.02 & 0.02 & 0.02 & 0.05\\
RMSE & [\%] & \multicolumn{1}{r}{12} & \multicolumn{1}{r}{13} & \multicolumn{1}{r}{23} & \multicolumn{1}{r}{22} & \multicolumn{1}{r}{15} & \multicolumn{1}{r}{11} & \multicolumn{1}{r}{29} & \multicolumn{1}{r}{11} & \multicolumn{1}{r}{14} & \multicolumn{1}{r}{ 6} & \multicolumn{1}{r}{12}\\
\bottomrule
\end{tabu}
\end{table}

\section{Conclusion}
\label{sec:conclusion}
We propose a fully automatic, AbS based method that permits the accurate spatial modeling of the thickness distribution of cortical bone from clinical QCT.
From a probabilistic bone model, stochastic measurement processes are synthesized.
The maximum a-posteriori model parameters, including cortical thickness, are estimated in an optimized MCEM process exploiting the different characteristics of the in-plane and out-of-plane PSF of clinical CT scanners.

We showed that our method is in tight agreement with the gold standard and completely eliminates the overestimation induced by other methods.
Besides from giving accurate thickness estimates per specimen, it permits the assessment of intra-bone thickness variations.
Because of its high accuracy our AbS method has the potential in improving estimation of bone strength.
The resulting spatial model of the cortical bone also has the potential in increasing the accuracy of FEA and osteoporosis diagnosis and monitoring in general.

There is of course a limitation:
the method was only validated on a few specimens, ex-vivo.
However, in in-vivo there is no gold standard available so accuracy cannot be analyzed directly.
The in-vivo validation could only be performed indirectly, e.g. by checking if fracture risk prediction or the distinction of treatment effects caused by different drugs can be improved.
This will be future work.

\subsubsection{Acknowledgments.}
This work was supported by the German Research Foundation, DFG, No. KO2044/9-1, 
and the Bundesministerium für Bildung (BMBF), Förderkennzeichen 01EC1005
(Diagnostik Bilanz Study, BioAsset Project).

\bibliography{paper784.bbl}

\begin{thebibliography}{10}
\providecommand{\url}[1]{\texttt{#1}}
\providecommand{\urlprefix}{URL }
\providecommand{\doi}[1]{https://doi.org/#1}

\bibitem{BUIE2007505}
Buie, H.R., Campbell, G.M., Klinck, R.J., MacNeil, J.A., Boyd, S.K.: Automatic
  segmentation of cortical and trabecular compartments based on a dual
  threshold technique for in vivo micro-ct bone analysis. Bone  \textbf{41}(4),
   505 -- 515 (2007). \doi{https://doi.org/10.1016/j.bone.2007.07.007}

\bibitem{caffo2005}
Caffo, B.S., Jank, W., Jones, G.L.: Ascent-based monte carlo expectation–
  maximization. Journal of the Royal Statistical Society: Series B (Statistical
  Methodology)  \textbf{67}(2),  235--251 (2005).
  \doi{10.1111/j.1467-9868.2005.00499.x}

\bibitem{DAMM2019194}
Damm, T., Peña, J.A., Campbell, G.M., Bastgen, J., Barkmann, R., Glüer, C.C.:
  Improved accuracy in the assessment of vertebral cortical thickness by
  quantitative computed tomography using the iterative convolution optimization
  (icon) method. Bone  \textbf{120},  194 -- 203 (2019).
  \doi{https://doi.org/10.1016/j.bone.2018.08.024}

\bibitem{ElLaham2018}
{El-Laham}, Y., {Elvira}, V., {Bugallo}, M.F.: Robust covariance adaptation in
  adaptive importance sampling. IEEE Signal Processing Letters  \textbf{25}(7),
   1049--1053 (July 2018). \doi{10.1109/LSP.2018.2841641}

\bibitem{ESWARAN2007}
Eswaran, S.K., Bayraktar, H.H., Adams, M.F., Gupta, A., Hoffmann, P.F., Lee,
  D.C., Papadopoulos, P., Keaveny, T.M.: The micro-mechanics of cortical shell
  removal in the human vertebral body. Computer Methods in Applied Mechanics
  and Engineering  \textbf{196}(31),  3025 -- 3032 (2007).
  \doi{https://doi.org/10.1016/j.cma.2006.06.017}

\bibitem{Hildebrand1997}
Hildebrand, T., Rüegsegger, P.: A new method for the model-independent
  assessment of thickness in three-dimensional images. Journal of Microscopy
  \textbf{185}(1),  67--75 (1997). \doi{10.1046/j.1365-2818.1997.1340694.x}

\bibitem{Kazakia2008}
Kazakia, G.J., Burghardt, A.J., Cheung, S., Majumdar, S.: Assessment of bone
  tissue mineralization by conventional x-ray microcomputed tomography:
  Comparison with synchrotron radiation microcomputed tomography and ash
  measurements. Medical Physics  \textbf{35}(7),  3170--3179 (2008).
  \doi{10.1118/1.2924210}

\bibitem{KAZAKIA2011}
Kazakia, G.J., Burghardt, A.J., Link, T.M., Majumdar, S.: Variations in
  morphological and biomechanical indices at the distal radius in subjects with
  identical bmd. Journal of Biomechanics  \textbf{44}(2),  257 -- 266 (2011).
  \doi{https://doi.org/10.1016/j.jbiomech.2010.10.010}

\bibitem{Krouwer2008}
Krouwer, J.S.: Why bland–altman plots should use x, not (y+x)/2 when x is a
  reference method. Statistics in Medicine  \textbf{27}(5),  778--780 (2008).
  \doi{10.1002/sim.3086}

\bibitem{laib1998}
Laib, A., H{\"a}uselmann, H.J., R{\"u}egsegger, P.: In vivo high resolution
  3d-qct of the human forearm. Technology and health care  \textbf{6}(5-6),
  329--337 (1998). \doi{10.3233/THC-1998-65-606}

\bibitem{laval1983}
Laval-Jeantet, A.M., Bergot, C., Carroll, R., Garcia-Schaefer, F.: Cortical
  bone senescence and mineral bone density of the humerus. Calcified tissue
  international  \textbf{35}(1),  268--272 (1983).
  \doi{https://doi.org/10.1007/BF02405044}

\bibitem{Lee2012}
{Lee}, P.M.: {Bayesian statistics. An introduction. 4th ed.} Chichester: John
  Wiley \& Sons, 4th ed. edn. (2012)

\bibitem{Li2015}
Li, C., Jin, D., Chen, C., Letuchy, E.M., Janz, K.F., Burns, T.L., Torner,
  J.C., Levy, S.M., Saha, P.K.: Automated cortical bone segmentation for
  multirow-detector ct imaging with validation and application to human
  studies. Medical Physics  \textbf{42}(8),  4553--4565 (2015).
  \doi{10.1118/1.4923753}

\bibitem{Liu2014}
{Liu}, Y., {Jin}, D., {Li}, C., {Janz}, K.F., {Burns}, T.L., {Torner}, J.C.,
  {Levy}, S.M., {Saha}, P.K.: A robust algorithm for thickness computation at
  low resolution and its application to in vivo trabecular bone ct imaging.
  IEEE Transactions on Biomedical Engineering  \textbf{61}(7),  2057--2069
  (2014). \doi{https://doi.org/10.1109/TBME.2014.2313564}

\bibitem{who2004}
Nojiri, S., Burge, R.T., Flynn, J.A., Foster, S.A., Sowa, H.: Who scientific
  group on the assessment of osteoporosis at primary health care level: summary
  meeting report. brussels, belgium (May 2004),
  \url{https://www.who.int/chp/topics/Osteoporosis.pdf}

\bibitem{Prevrhal1999}
Prevrhal, S., Engelke, K., Kalender, W.A.: Accuracy limits for the
  determination of cortical width and density: the influence of object size and
  {CT} imaging parameters. Physics in Medicine and Biology  \textbf{44}(3),
  751--764 (jan 1999). \doi{10.1088/0031-9155/44/3/017}

\bibitem{Prevrhal2003}
Prevrhal, S., Fox, J.C., Shepherd, J.A., Genant, H.K.: Accuracy of ct-based
  thickness measurement of thin structures: Modeling of limited spatial
  resolution in all three dimensions. Medical Physics  \textbf{30}(1), ~1--8
  (2003). \doi{10.1118/1.1521940}

\bibitem{R}
{R Core Team}: R: A Language and Environment for Statistical Computing. R
  Foundation for Statistical Computing, Vienna, Austria (2013),
  \url{http://www.R-project.org/}

\bibitem{Reinhold2019}
Reinhold, S., Damm, T., Huber, L., Andresen, R., Barkmann, R., Gl{\"u}er, C.C.,
  Koch, R.: An analysis by synthesis approach for automatic vertebral shape
  identification in clinical qct. In: Brox, T., Bruhn, A., Fritz, M. (eds.)
  Pattern Recognition. pp. 73--88. Springer International Publishing, Cham
  (2019). \doi{https://doi.org/10.1007/978-3-030-12939-2\_6}

\bibitem{Ritzel1997}
Ritzel, H., Amling, M., Pösl, M., Hahn, M., Delling, G.: The thickness of
  human vertebral cortical bone and its changes in aging and osteoporosis: A
  histomorphometric analysis of the complete spinal column from thirty-seven
  autopsy specimens. Journal of Bone and Mineral Research  \textbf{12}(1),
  89--95 (1997). \doi{10.1359/jbmr.1997.12.1.89}

\bibitem{smith2004treatment}
Smith, M., Dunkow, P., Lang, D.: Treatment of osteoporosis: missed
  opportunities in the hospital fracture clinic. Annals of the Royal College of
  Surgeons of England  \textbf{86}(5), ~344 (2004).
  \doi{https://doi.org/10.2298/VSP1205420D}

\bibitem{TREECE2015249}
Treece, G., Gee, A.: Independent measurement of femoral cortical thickness and
  cortical bone density using clinical ct. Medical Image Analysis
  \textbf{20}(1),  249 -- 264 (2015).
  \doi{https://doi.org/10.1016/j.media.2014.11.012}

\bibitem{TREECE2010276}
Treece, G., Gee, A., Mayhew, P., Poole, K.: High resolution cortical bone
  thickness measurement from clinical ct data. Medical Image Analysis
  \textbf{14}(3),  276 -- 290 (2010).
  \doi{https://doi.org/10.1016/j.media.2010.01.003}

\bibitem{TREECE2012952}
Treece, G., Poole, K., Gee, A.: Imaging the femoral cortex: Thickness, density
  and mass from clinical ct. Medical Image Analysis  \textbf{16}(5),  952 --
  965 (2012). \doi{https://doi.org/10.1016/j.media.2012.02.008}

\bibitem{YAMADA2019}
Yamada, S., Chiba, K., Okazaki, N., Era, M., Nishino, Y., Yokota, K., Yonekura,
  A., Tomita, M., Tsurumoto, T., Osaki, M.: Correlation between vertebral bone
  microstructure and estimated strength in elderly women: An ex-vivo hr-pqct
  study of cadaveric spine. Bone  \textbf{120},  459 -- 464 (2019).
  \doi{https://doi.org/10.1016/j.bone.2018.12.005}

\end{thebibliography}

\end{document}